\documentclass[lettersize,journal]{IEEEtran}
\usepackage{amsmath,amsfonts}
\usepackage{algorithmic}
\usepackage{array}
\usepackage{orcidlink}
\usepackage[caption=false,font=normalsize,labelfont=sf,textfont=sf]{subfig}
\usepackage{textcomp}
\usepackage{stfloats}
\usepackage{url}
\usepackage{verbatim}
\usepackage{graphicx}
\usepackage{pifont}
\usepackage{cite}
\usepackage{float}
\usepackage{xcolor}
\def\BibTeX{{\rm B\kern-.05em{\sc i\kern-.025em b}\kern-.08em
    T\kern-.1667em\lower.7ex\hbox{E}\kern-.125emX}}
\usepackage{balance}

\newcommand\copyrighttext{%
  \footnotesize \textcopyright 2022 IEEE. Personal use of this material is permitted.
  Permission from IEEE must be obtained for all other uses, in any current or future
  media, including reprinting/republishing this material for advertising or promotional
  purposes, creating new collective works, for resale or redistribution to servers or
  lists, or reuse of any copyrighted component of this work in other works.
  DOI: \href{<http://tex.stackexchange.com>}{10.1109/TIM.2022.3218303}}
\newcommand\copyrightnotice{%
\begin{tikzpicture}[remember picture,overlay]
\node[anchor=south,yshift=10pt] at (current page.south) {\fbox{\parbox{\dimexpr\textwidth-\fboxsep-\fboxrule\relax}{\copyrighttext}}};
\end{tikzpicture}%
}

\begin{document}
\title{Augmented Reality and Mixed Reality Measurement Under Different Environments: A Survey on Head-Mounted Devices}
\author{Hung-Jui~Guo\orcidlink{0000-0003-2233-846X},
        Jonathan Z. Bakdash\orcidlink{0000-0002-1409-4779},
        Laura R. Marusich\orcidlink{0000-0002-3524-6110},
        and~Balakrishnan~Prabhakaran\orcidlink{0000-0003-0385-8662},~\IEEEmembership{Senior Member,~IEEE}
\thanks{Manuscript received April 19, 2022; revised September 1, 2022; accepted October 12, 2022. This research was sponsored by the DEVCOM U.S. Army Research Laboratory under Cooperative Agreement Number W911NF-21-2-0145 to B.P.}
\thanks{Authors' addresses: Hung-Jui Guo, hxg190003@utdallas.edu, The University of Texas at Dallas, 800 W Campbell Rd, Richardson, Texas, USA, 75080; Jonathan Z. Bakdash, jonathan.z.bakdash.civ@army.mil, jbakdash@gmail.com, U.S. Army Combat Capabilities Development Command Army Research Laboratory South at the University of Texas at Dallas, 800 W Campbell Rd, Richardson, Texas, USA, 75080 and Department of Psychology and Special Education, Texas A\&M University-Commerce, 2200 Campbell St, Commerce, Texas, USA, 75428; Laura R. Marusich, laura.m.cooper20.civ@army.mil, U.S. Army Combat Capabilities Development Command Army Research Laboratory South at the University of Texas at Arlington, 701 S Nedderman Dr, Arlington, Texas, USA, 76019; Balakrishnan Prabhakaran, bprabhakaran@utdallas.edu, The University of Texas at Dallas, 800 W Campbell Rd, Richardson, Texas, USA, 75080.}
}

\markboth{Journal of \LaTeX\ Class Files,~Vol.~18, No.~9, September~2020}{Augmented Reality and Mixed Reality Measurement Under Different Environments: A Survey on Head-Mounted Devices}

\maketitle
\copyrightnotice

\begin{abstract}
Augmented Reality (AR) and Mixed Reality (MR) have been two of the most explosive research topics in the last few years. Head-Mounted Devices (HMDs) are essential intermediums for using AR and MR technology, playing an important role in the research progress in these two areas. Behavioral research with users is one way of evaluating the technical progress and effectiveness of HMDs. In addition, AR and MR technology is dependent upon virtual interactions with the real environment. Thus, conditions in real environments can be a significant factor for AR and MR measurements with users. 
In this paper, we survey 87 environmental-related HMD papers with measurements from users, spanning over 32 years. We provide a thorough review of AR- and MR-related user experiments with HMDs under different environmental factors. Then, we summarize trends in this literature over time using a new classification method with four environmental factors, the presence or absence of user feedback in behavioral experiments, and ten main categories to subdivide these papers (e.g., domain and method of user assessment). We also categorize characteristics of the behavioral experiments, showing similarities and differences among papers. 
\end{abstract}


\section{Introduction}
\IEEEPARstart{C}{urrently} there is widespread and increasing interest in using virtual worlds and virtual objects for entertainment, education, training, and research applications. This includes the metaverse, a single fully-connected virtual universe. The most well known and widely used technologies for virtual environments are Virtual Reality (VR), Augmented Reality (AR),  and Mixed Reality (MR).  
In AR and MR specifically, the research focus is on the integration of virtual objects with real environments across a variety of applications and aspects of daily human life.

Because of the wide and growing number of potential applications, it is likely that AR and MR will have greatly increased market value in the near future. According to the Augmented Reality \& Mixed Reality Market Trend article in ReportLinker \cite{ARMRTrend2021}, AR/VR demand has increased significantly due to COVID-19. Moreover, AR's market value growth rate is expected to reach 83.3\% from its current value, while the growth rate of MR is expected to be 41.8\% from its current value, which means that these fields will have a high degree of development in the next few years.
The Google Books Ngram Viewer \cite{GoogldBooksNgramViewer} shows that AR increased significantly in the book mentions proportion around 1990 and grew to about 300 times in 2019. Similarly, MR increased significantly in mentions proportion around 1994 and grew by a factor of approximately 46 times in 2019.

In recent years, industry and academic researchers have developed numerous AR, VR, and MR Head-Mounted Devices (HMDs), which combine virtual information with information in real-world environments. A user wearing an HMD enters a virtual or real-virtual hybrid world, enabling interactions that are not generally possible in the physical world alone. This capability has led to the use of HMDs in many different fields to introduce and develop new interactive methods. Example applications include entertainment/gaming (e.g. \cite{Quest2Web}), training (e.g., \cite{ZHANG2017717}), education (e.g., \cite{Tang2020}), healthcare (e.g., \cite{Wittich2018}), and design (e.g., \cite{Vitali2018}).  

A large body of existing research and development of HMDs focuses on the two primary goals of making users feel at ease in virtual environments and at an affordable price. Particularly for the first goal, assessments of human users in the HMDs is indispensable to developing the devices and their corresponding software. Similarly, researchers in the various application fields described above (entertainment, education, etc.) rely on human studies to guide the development of  virtual object display techniques and user interaction methods. In this survey paper, we summarize the literature for research with HMDs and user measurement using different categories. Our main target was papers assessing AR and/or MR HMDs, but because many measurement papers use VR HMDs or use both VR and AR devices in the same experiments, we also included papers using VR HMDs.
 
This paper provides three new contributions over existing related literature surveys:
\begin{itemize}
\item \textbf{Latest (as of early 2022) environmental-related and HMD-related measurement survey paper}: We summarize the AR/MR HMD literature from 1990 to early 2022, including trends over time, identifying gaps for future research directions.
\item \textbf{Establish four environmental factor categories and ten new measurement categories}: We have integrated the environmental factors commonly used in existing papers and divided them into four main categories. We further subdivided these papers into ten main categories of measurement.
\item \textbf{Unify different measurement experiment designs}: We analyzed different user measurement papers and summarized the experiment designs, including participant attributes and experimental procedure.
\end{itemize}

\section{Background}
Here we define and summarize the relevant background literature about the key elements addressed in this survey paper: AR, MR, virtual environments, and HMDs.

\subsection{Augmented Reality}
In AR, users see virtual objects (usually displayed through a handheld device) appearing in the real-world environment. In the AR environment, the virtual objects seen by the users are rendered virtually.
Common examples are Nintendo's Pokémon Go App and the IKEA Mobile App.

The first concept related to AR appeared in 1901 in the novel \textit{The Master Key}, written by Lyman Frank Baum, which describes a child who experimented with a wearable electronic display to change and overlay data onto a real-world environment. This concept was later implemented by researchers and has became increasingly popular. Many AR-related survey papers have been written to discuss existing AR capabilities, applications, and limitations with technology. Moreover, according to \cite{Van10}, a description of the world's first HMD using AR was published in 1960 \cite{Tamura12}.

The first and currently most cited AR-related survey paper was published in 1997 by Dr. Azuma \cite{Azuma97}, focusing on categories of applications for AR and registration problems. A subsequent (2001) highly-cited survey paper \cite{Azuma01} described methods of displaying AR applications such as "Head-worn displays" (i.e., HMDs), handheld displays, projection displays, as well as challenges for AR displays, such as outdoor environments and positional sensing.  
Then, in 2010, \cite{Van10} published a survey paper for the AR-related research linking AR and MR. In this paper, AR and MR were linked using the concept of the reality-virtuality continuum. Similar to \cite{Azuma01}, this article summarized the display options and application examples for HMDs and also promising new technologies that have made rapid recent progress, such as optical see-through \cite{Gudrun01} and spatial displays.

\subsection{Mixed Reality}
Similar to AR, MR also creates virtual objects in a real-world environment using the environment for hybrid rendering methods to generate objects. 
A current common example is the Microsoft HoloLens \cite{hololensWeb}.
The most significant difference between AR and MR is that the interaction between virtual items and the real environment is added to the MR environment. This difference allows the virtual objects in the MR environment to better simulate and correspond to the physical conditions of the real-world environment. Users can use their own hands to interact with virtual objects directly.

MR was first introduced in 1994 \cite{mixedRealityOriginal, Milgram95}, extending AR to see-through and monitor-based displays. These authors used a continuum  to present the relation between the real environment, AR, Augmented Virtuality (AV), and the virtual environment.
A more complete concept of MR was proposed in \cite{Benford94}, which is an extension and mixing of AR and VR. This paper included the concept of space, objects, and users' activity, which established an important foundation for future MR-related work.
A further extension \cite{Benford1998} introduced more detailed concepts and specific applications, as well as the possibility of collaboration (i.e., multiple simultaneous users) in MR. Collaboration in MR has since been highly used in research and applications (e.g., Microsoft Mesh: \url{https://www.microsoft.com/en-us/mesh}). 
Similarly, \cite{Billinghurst99} presented collaborative MR works and their own calibration method. 

There are also papers describing the concept of MR \emph{combined} with AR and VR \cite{Mann18, Bekele18, Speicher19}. In addition, there are multiple MR papers focusing on application domains such as education \cite{Hughes05, Pan06, Tang20} and entertainment \cite{Fiorentino02, Cheok09}. 

\subsection{Virtual Environment}

Virtual environments are widely used in various research fields such as computer science and psychology. In \cite{Liu14}, the authors focused on distributed virtual environments, which incorporate Internet communication and protocols. 
Another group targeted potential uses of brain-computer interfaces in interactive virtual environments \cite{Kerous16}. 
A recent survey paper on presence in virtual environments \cite{Souza21} suggests that the sense of presence has a positive correlation with authenticity \cite{Sjolie13}. 

In addition to review papers, many individual papers describe research on users in virtual environments.  For example, \cite{Fisher87} discussed the display system related to the virtual environment, providing an HMD prototype of the virtual environment display system, including hand position tracking sensors. 

Later, \cite{Mine95} introduced interaction techniques in the virtual environment, including users' movement, selection techniques, manipulation (position/rotation), and scaling method. Although this paper was published in 1995, their implementation was presciently close to technology in 2022. 
In a different approach, \cite{Chen95}, mainly focused on cameras (rotation and zooming) for image-based virtual environments using modeling and rendering. 

To evaluate interaction techniques in virtual environments, \cite{Bowman01} proposed a testbed evaluation method that provided multiple performance metrics for interactions, including object selection and manipulation as well as travel. 
Lastly, in \cite{Blascovich02}, Immersive Virtual Environment Technology (IVET) used virtual avatars for research in social psychology such as self and group identity.  

\begin{figure*}[h]
\centering
\subfloat[]{\includegraphics[width=0.15\textwidth, keepaspectratio=true]{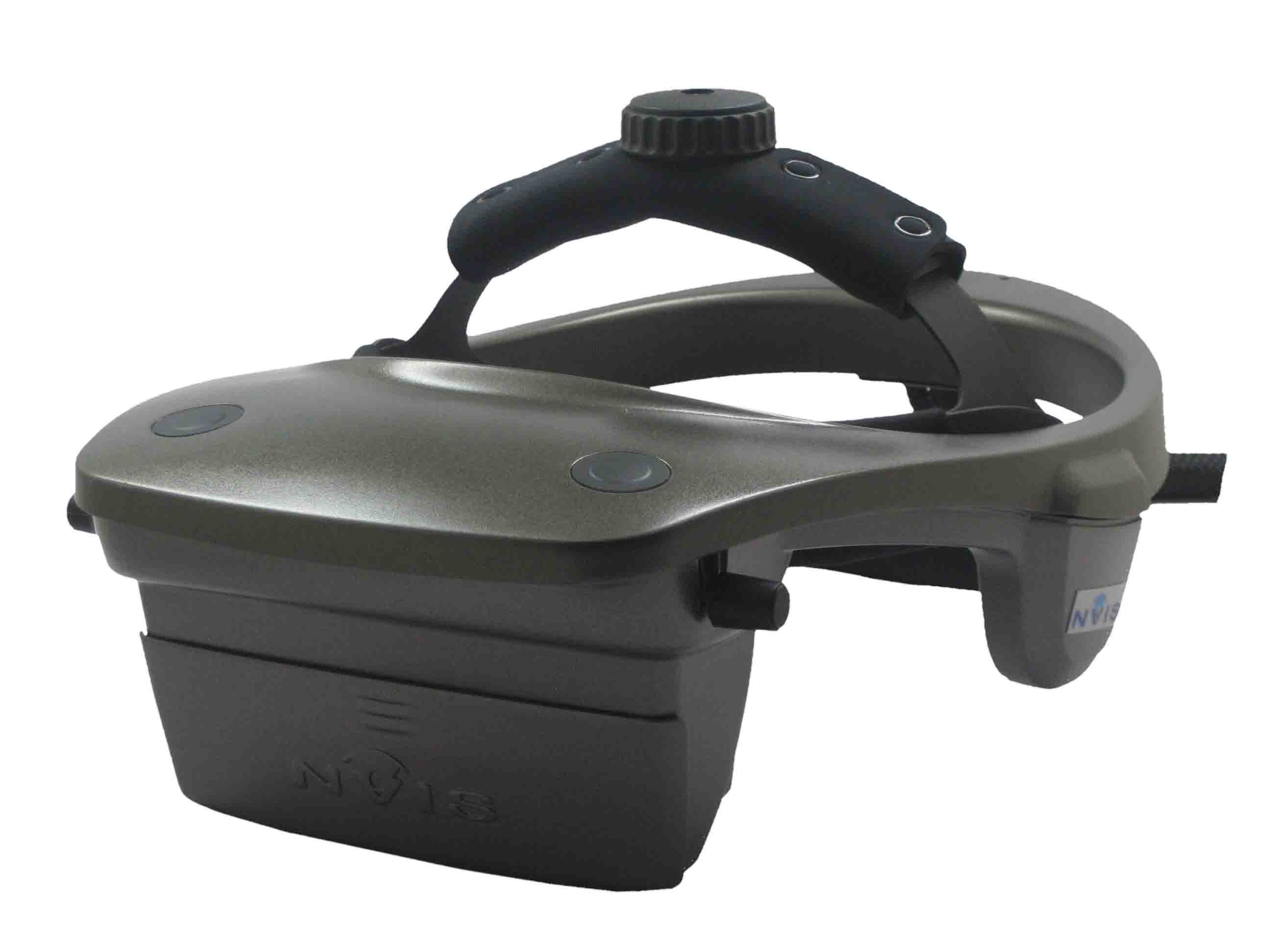}%
\label{fig:nVisor}}
\hfil
\subfloat[]{\includegraphics[width=0.15\textwidth, keepaspectratio=true]{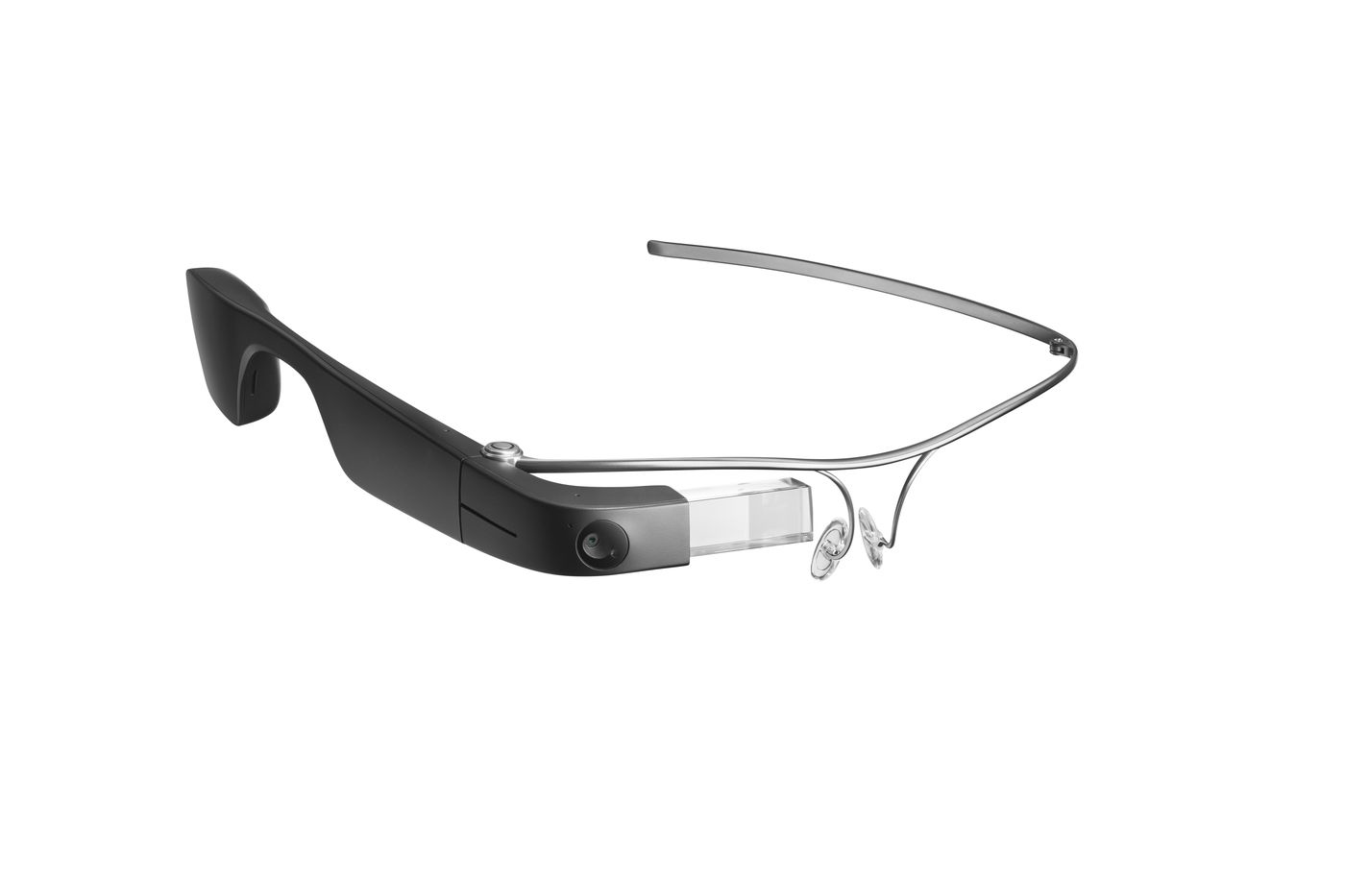}%
\label{fig:GoogleGlass}}
\hfil
\subfloat[]{\includegraphics[width=0.15\textwidth, keepaspectratio=true]{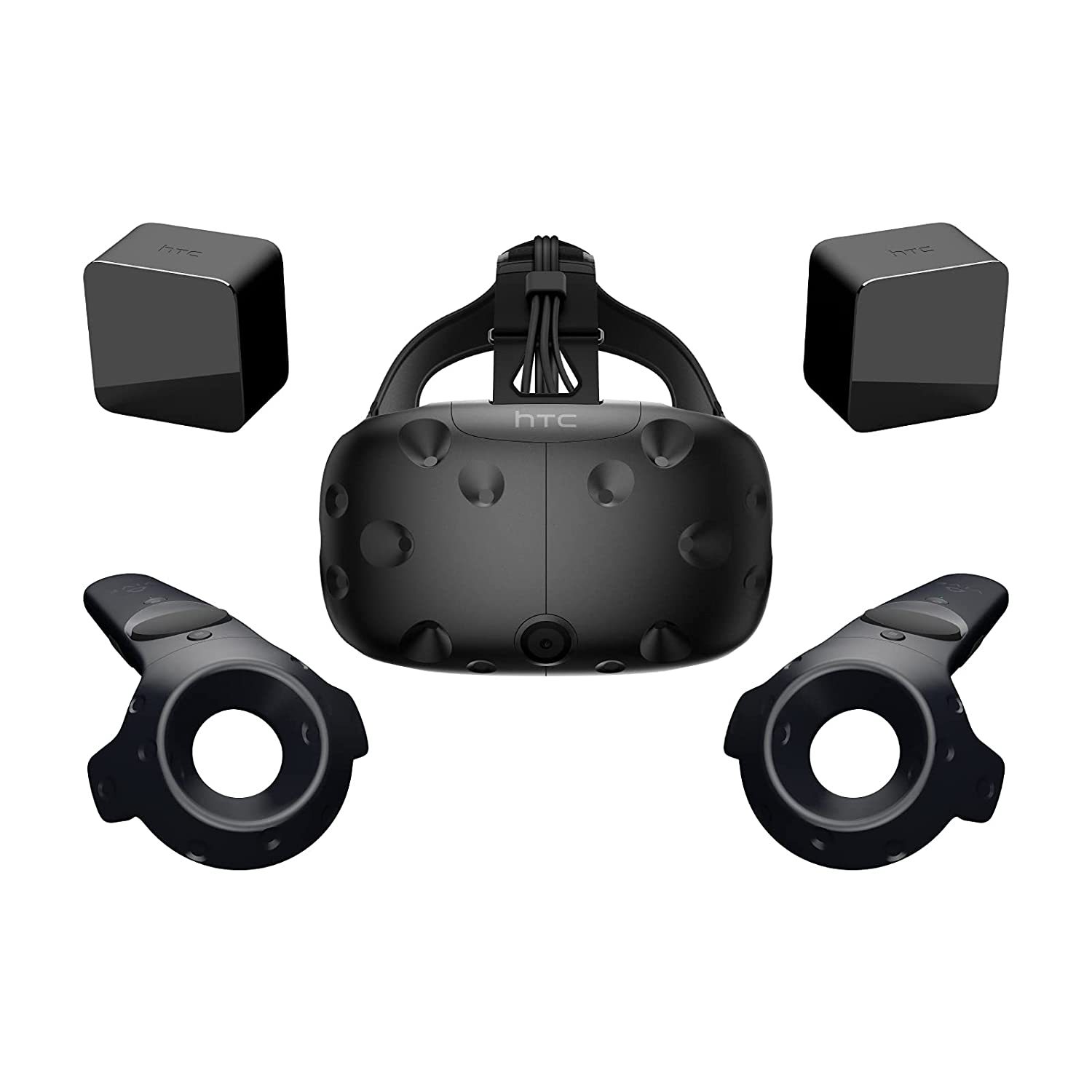}%
\label{fig:HTCVive}}
\hfil
\subfloat[]{\includegraphics[width=0.15\textwidth, keepaspectratio=true]{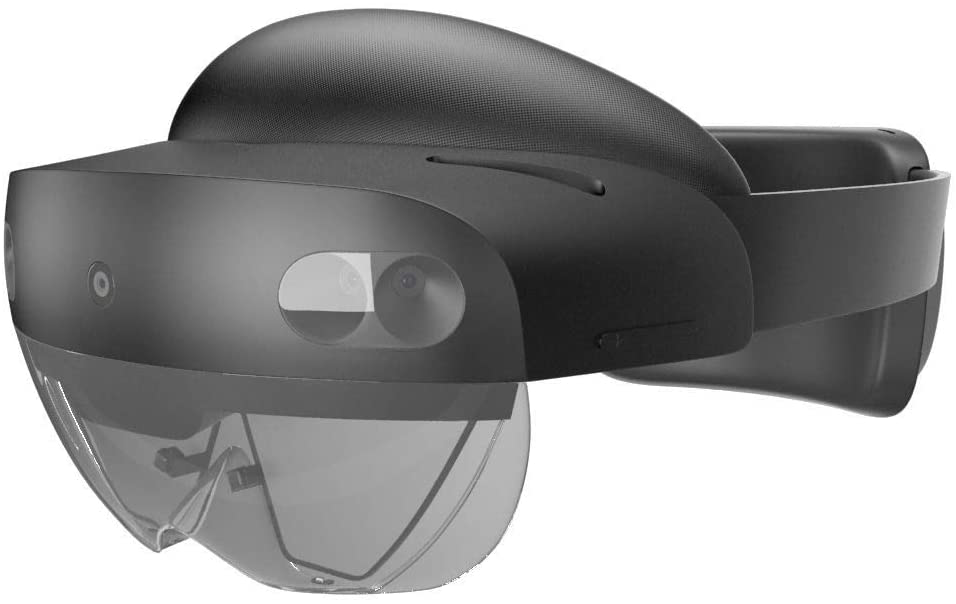}%
\label{fig:HL2}}
\hfil
\subfloat[]{\includegraphics[width=0.15\textwidth, keepaspectratio=true]{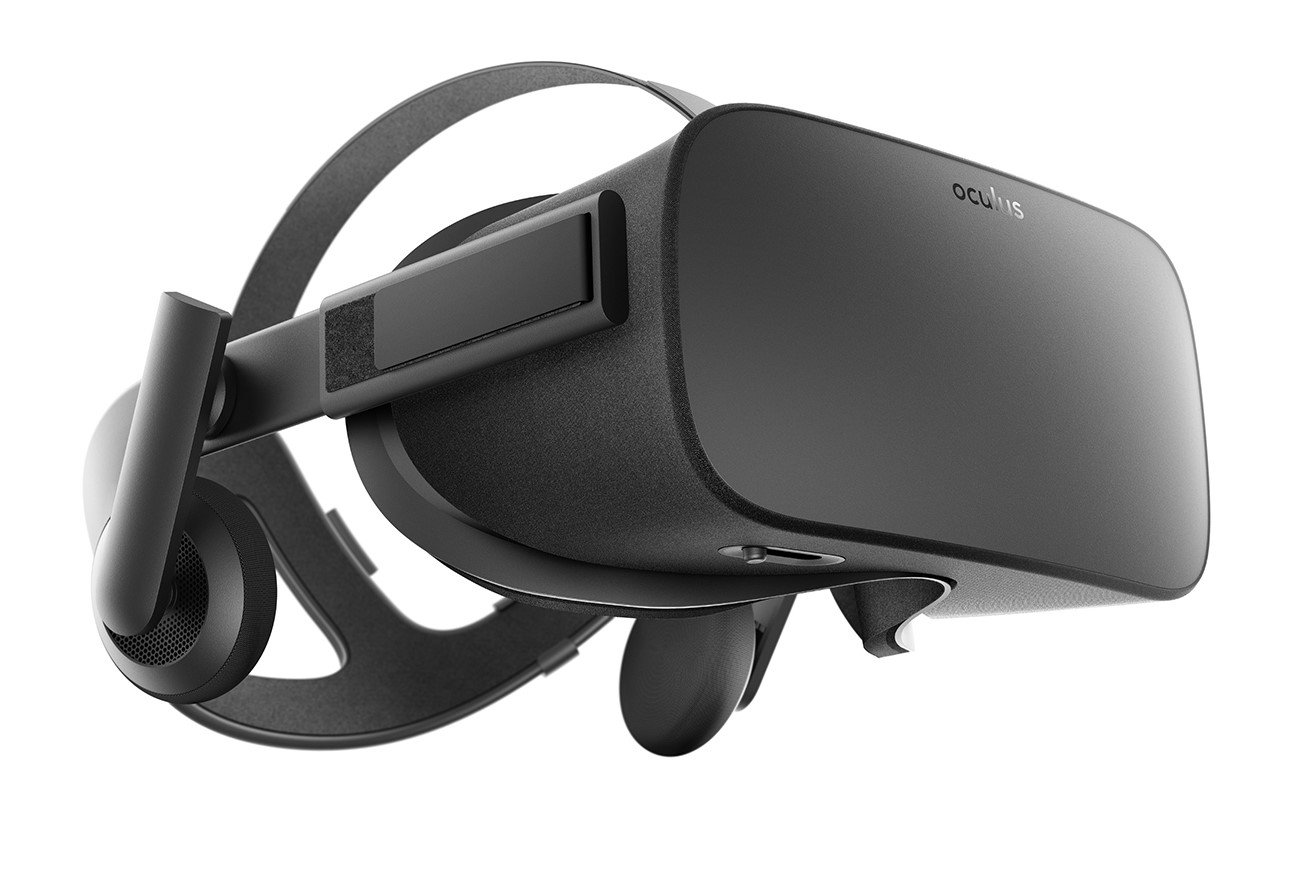}%
\label{fig:OculusRift}}
\hfil
\subfloat[]{\includegraphics[width=0.15\textwidth, keepaspectratio=true]{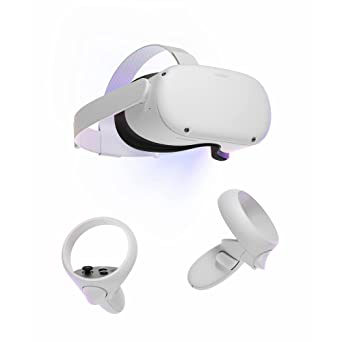}%
\label{fig:Quest2}}
\caption{Widely used Head-Mounted Devices (HMDs). (a) NVIS nVisor HMD \cite{NVISWeb}. (b) Google Glass HMD \cite{GoogleGlassWeb}. (c) HTC Vive HMD  \cite{HTCViveWeb}. (d) Microsoft HoloLens 2 HMD \cite{hololensWeb}. (e) Oculus Rift HMD \cite{RiftWeb}. (f) Oculus Quest 2 HMD \cite{Quest2Web}}
\label{fig:HMDs}
\vspace{-1em}
\end{figure*}

\subsection{Head-Mounted Devices}
As described earlier, one of the most widely known tools for using AR, VR, and MR is the HMD, a helmet-shaped device that can move with the user and has screens in front of one or both eyes to display objects in the virtual environment. As of early 2022, the most commonly used HMDs were the NVIS nVisor \cite{NVISWeb}, Google Glass \cite{GoogleGlassWeb}, HTC Vive \cite{HTCViveWeb}, Microsoft HoloLens \cite{hololensWeb}, Oculus Rift \cite{RiftWeb}, and Oculus Quest \cite{Quest2Web}, pictured in Figure \ref{fig:HMDs}.   

One category of HMD is the Optical See-Through Head-Mounted Device (OST-HMD). This technology uses optical methods to reflect virtual objects and allows the real-world objects to pass through the lens in front of the user's eyes to achieve the appearance of virtual objects and real-world objects together in the same space. OST-HMDs are particularly useful for AR and MR.

In 1968, Dr. Sutherland published one of the earliest papers \cite{Sutherland68} regarding a complete design of an HMD. This paper introduced the display system for the HMD and the head position sensor for tracking.
Later work \cite{Silva03} described hardware architecture for AR devices, including OST-HMD, virtual retinal system HMD, and video see-through HMD.
In a comparison of user navigation performance, \cite{Santos09} found that  performance was better with a desktop display than an HMD, although most users were satisfied with the HMD.   

There have also been several survey papers reviewing different research areas using HMDs, usually associated with AR and VR. 
Some review papers have focused on HMD hardware. For example, \cite{Rolland05} provided a survey of HMD history such as Active-Matrix Liquid-Crystal-Displays (AM-LCDs), Ferroelectric Liquid Crystal on Silicon (FLCOS) \cite{Wu01}, Organic Light Emitting Displays (OLEDs) \cite{Rajeswaran00}, and Time Multiplex Optical Shutter (TMOS) \cite{Selbrede06}. 

There have also been comparisons among different HMDs in user experiments. In \cite{Buck18}, the authors compared the distance estimation in multiple HMDs such as Oculus Rift, Nvis, and HTC Vive.
Returning to HMD hardware, \cite{Itoh21} published a survey paper on state-of-the-art OST-HMD-related techniques for AR and MR, such as Microsoft HoloLens and HoloLens 2 \cite{hololensWeb}. 
This survey focused on visual coherence (i.e., blending of virtual and real content) in OST-HMDs in AR and MR research areas. The authors also discussed existing challenges among OST-HMD-related papers, divided into three key areas: spatial realism, temporal realism, and visual realism.

There are a variety of examples of specific applications for HMDs. For example, \cite{Profita16} discussed the social acceptability of disabled/non-disabled people wearing HMDs. 
\cite{Cubelos19} analyzed the perception of view transition for dense multi-view content when users are wearing HMDs and transitioning their views.
In \cite{Culham04}, the authors compared four different kinds of HMDs with conventional optical low-vision aids.
Then, \cite{Cummiskey17} used HMDs to investigate the causal agents of head trauma in athletes.
Also, in \cite{Lin22}, the authors used HMDs to compare user preferences for different virtual navigation instructions that integrate with the real environment in the MR world. 
Lastly, \cite{Arefin22} built a custom HMD to evaluate performance and eye fatigue for context and focal switching in AR.

The above background knowledge demonstrates how interconnected the topics of AR, MR, virtual environments, and HMDs are. Also, many measurement-related papers use the tools mentioned in these topics for experiments. Therefore, we aim to integrate these topics, and organize and discuss papers that use HMDs and environmental factors as the primary measurement goals.

\section{Literature Review}
To focus our literature review, we searched for the terms "HMD" with "measurement" and "environmental factors" in various combinations in ACM Digital Library, IEEE Xplore, SPIE Digital Library, Springer Link, and other libraries. We also used Google Scholar to find additional papers published on other platforms or otherwise missed. 
After analyzing the content of papers found using the above keywords, we used our own defined paper categories to add keywords in advanced searches---for example: "HMD signal measurement in environment" or "HMD comfort assessment in virtual environment."

We targeted measurement papers from 1990 to early 2022 and required that the content of included papers must have three of the following four types of information: environmental factors, participant information (e.g., Age distribution, Male/Female ratio), data collection method, and the type of HMD(s) used in the experiment.

We also collected other information from these papers, such as measurement method, experimental procedure, data processing method, and challenges/limitations encountered, and organized them in the following sections. Categorizing and summarizing these papers by their methodology allows us to identify gaps in the literature, and inform methods for future experiments.

\begin{figure*}[h]
\centering
\includegraphics[width=0.8\textwidth, keepaspectratio=true]{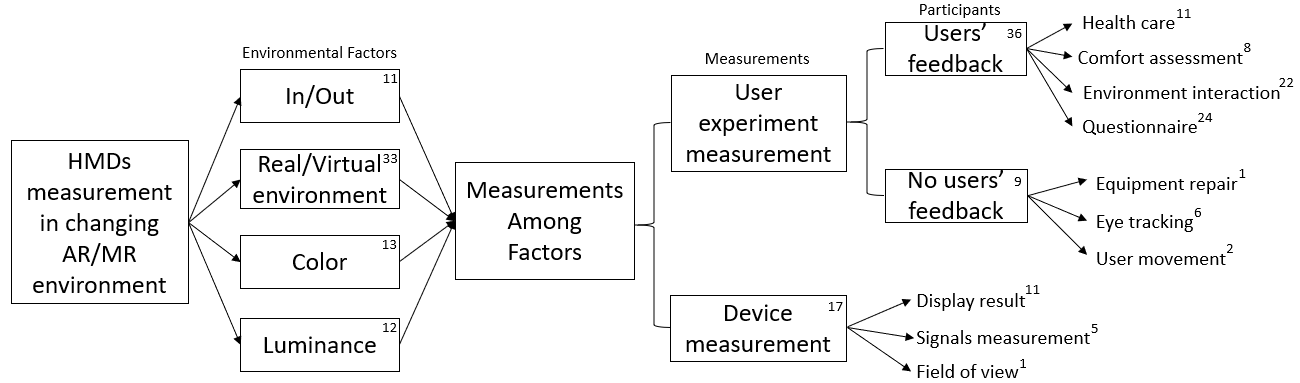}
\caption{Categories we used to classify related works in this paper. The numbers in the upper right corner of each category represent the number of related works in each category.}
\label{fig:Categories}
\vspace{-1em}
\end{figure*}

\section{Categorization: Environmental Factors, User or Device Measurements, and Participants}

In this section, we introduce our method of categorizing the relevant papers. This is visualized in the tree diagram in Figure \ref{fig:Categories}. In this diagram, the number of measurement-related papers assigned to each category appears in the upper right corner of that category (some papers may be classified in more than one category). First, we divide  papers into four categories according to the main environmental factors manipulated and/or assessed in the papers, which are: indoor environment vs. outdoor environment (In/Out) \cite{Nakao14, Renkewitz2008, DeFelice2014, {Rangelova2018}}, virtual environment vs. virtual/real environment (Real/Virtual environment) \cite{Robert2016, Patrick2000}, color difference \cite{Shin2020, Do20}, and luminance difference \cite{Birkfellner2002, Cognolato2018}. (The reference list for each environmental factor is listed in the Appendix \ref{FirstAppendix})  
We provide more detailed information about these four environmental factor categories in section \ref{EnvironmentalFactorsSection}.

We further divided the related papers according to the type of measurement conducted: user experiment measurement or device measurement \cite{Ryu2017, Chan2019}, as shown in the right half of Figure \ref{fig:Categories}. The primary classification basis for these two categories is whether the researcher assesses some aspect of human experience or behavior (e.g., action patterns, reactions, or feedback), or assesses only aspects of the HMD.

User experiment measurement studies may be divided into two additional categories: users' feedback and no users' feedback. This classification is based on whether the experiment requires any type of response or input behavior from participants. Examples of users' feedback include asking participants to report a target's color \cite{Shin2020,Zhang2021}, to fill in a questionnaire after the experiment is over \cite{Shu19,Wittich2018}, or automatically logged responses, such as time spent on achieving pre-set goals \cite{Kahl21,Lin22}, or accuracy in putting virtual objects in the correct position in the environment \cite{Swan06,Khan21}.
Studies that incorporate users' responses are the most common category among the types of measurement, likely because capturing direct feedback from users is a quick and easy way to obtain experimental data. Frequently used cases in this category are: health care (participants report their symptoms after using HMDs) \cite{MonWilliams1993, Peli1998}, comfort assessment (participants report whether they feel uncomfortable after using HMDs in different environments) \cite{Ha2020, Wille2013}, environment interaction (participants provide feedback on changes in the environment) \cite{Peer2017, Erickson20}, and questionnaire (the participants are asked to fill in questionnaires before or after the experiment) \cite{Profita16, Wang2019}.

In the less common category of no users' feedback, experimenters observe participants' actions externally and record them as experimental data. Instead of asking for specific feedback or responses from participants, experimenters record (manually or with external measurement devices) participants' actions for comparison. Common use cases include equipment repair (participants to repair some instruments and experimenters observe their actions) \cite{Ariansyah2022}, eye tracking (experimenters record the movement of the eye by the machine and aggregate it into data) \cite{Imaoka2020, Lidegaard14}, and user movement (experimenters observe participants movement as they complete a specific goal) \cite{Mohler2007, Batmaz2019}.
Although there are relatively few published papers at present in the category of using HMDs for equipment repair, this topic has boomed in recent years, with technologists trying to use MR or AR devices (such as Microsoft HoloLens) to accomplish this goal. We placed equipment repair into a separate category, as we expect this area to continue to grow.

For studies that fall into the category of device (as opposed to user) measurement, the primary goal is typically to test the technical performance of HMDs, such as performance in different environments, HMDs' display quality, or the efficacy of HMD systems. Common themes of these papers are: display result \cite{Mine1993, Watson1997}, signals measurement \cite{Alamri2010, Weber2021}, and field of view measurement (FOV) \cite{Wittich2018}. Although these studies may require human participants to manipulate the devices, the main goal and focus of assessment is on quantifying technical aspects of HMDs.
(The reference list for each category is listed in the Appendix \ref{SecondAppendix})

In the next sections, we provide more detail on the papers included in the categories described above. Section \ref{EnvironmentalFactorsSection} discusses the environmental factors on the left half of Figure \ref{fig:Categories}, Section \ref{ExistingExperiments} covers participants' composition, data collection methods, and the mechanism (HMDs) used in related papers, and Section \ref{ExperimentalProcedureSection} covers the detailed structure of the experiment, corresponding to the right half of Figure \ref{fig:Categories}. 
Lastly, we provide general discussion, future directions and conclusions in Sections \ref{ChallengesSection} and \ref{ConclusionSection}.

\begin{figure*}[h]
\centering
\subfloat[]{\includegraphics[width=0.5\textwidth, keepaspectratio=true]{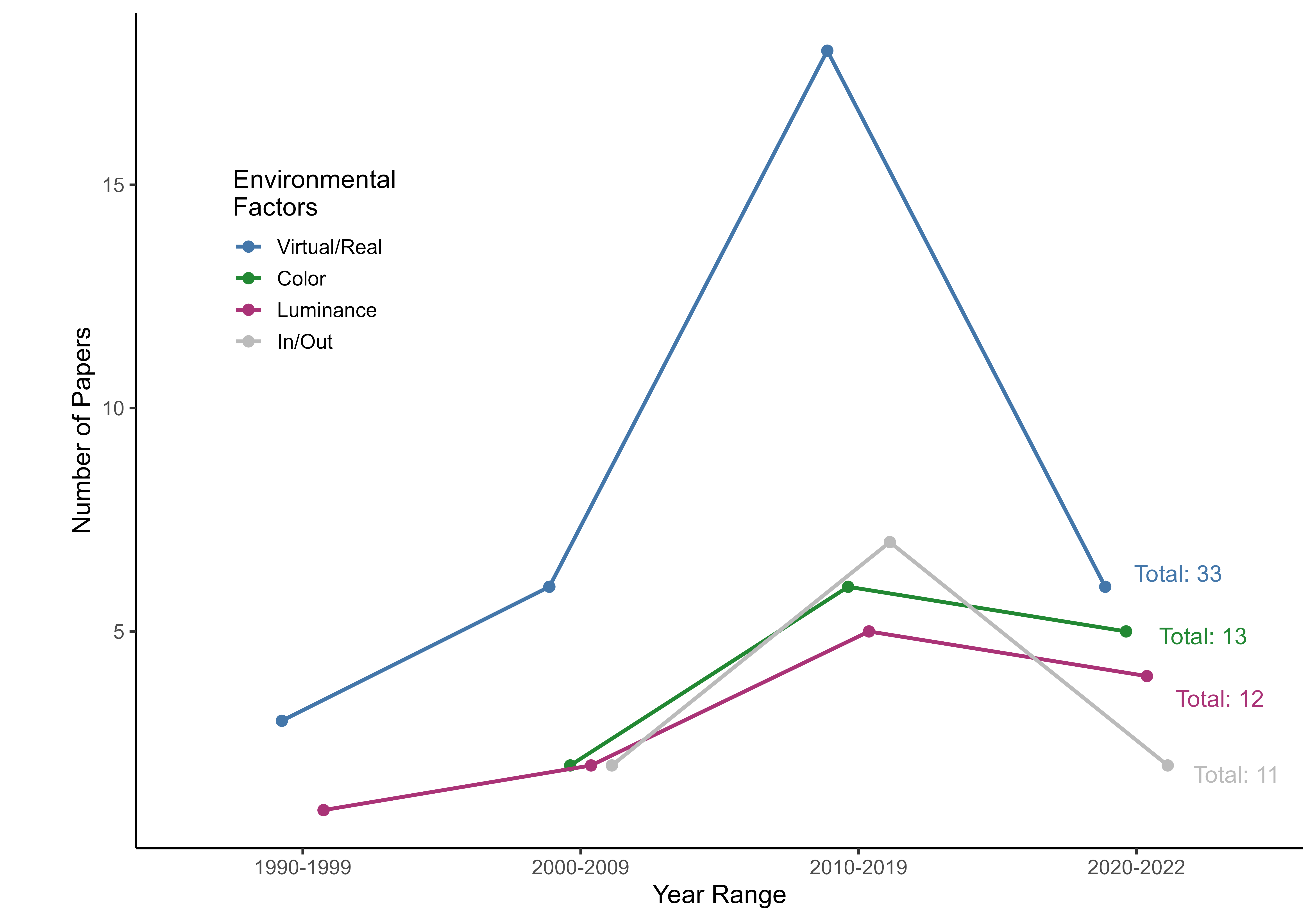}%
\label{fig:FactorsYear}}
\hfil
\subfloat[]{\includegraphics[width=.5\textwidth, keepaspectratio=true]{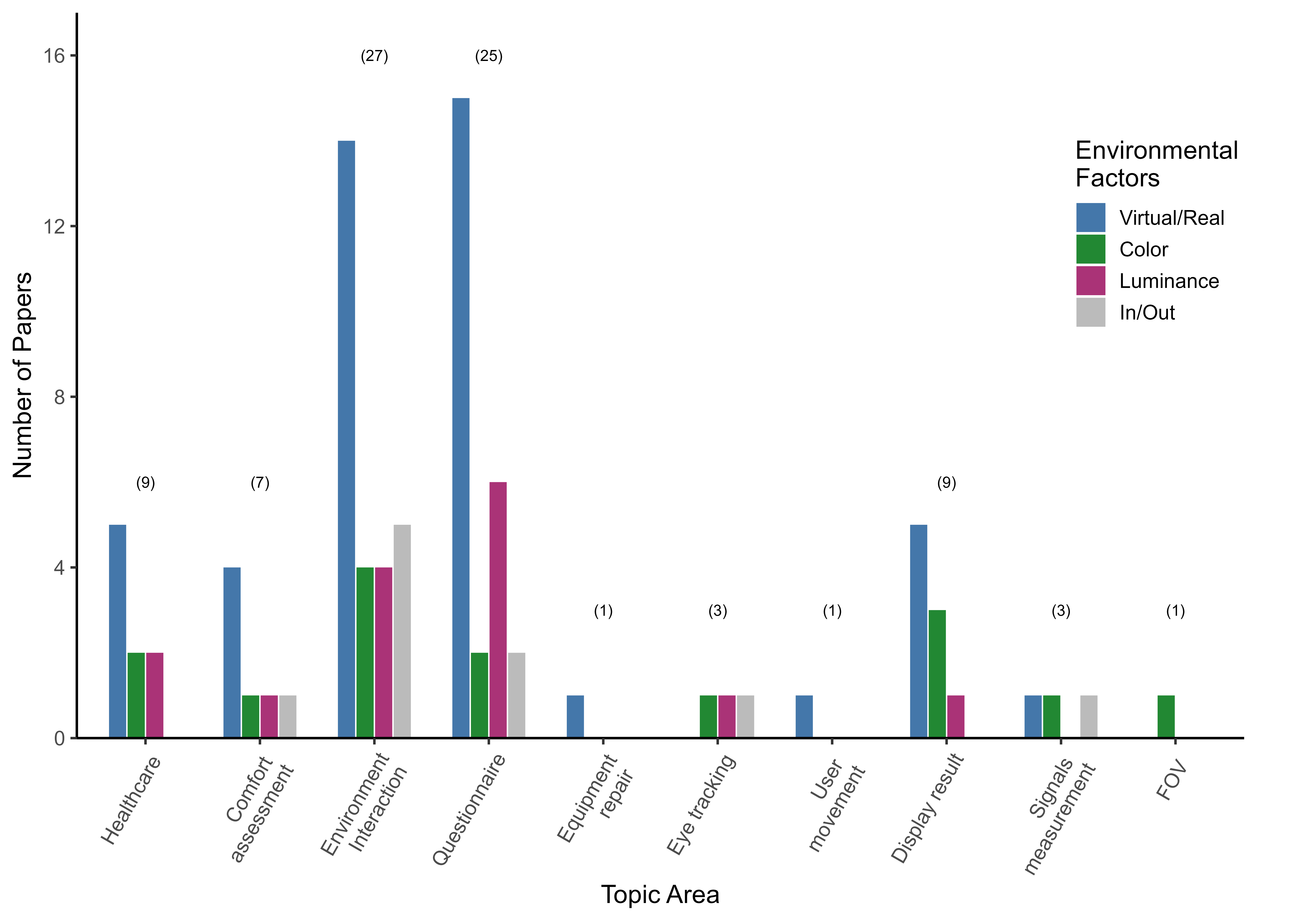}%
\label{fig:FactorsSummary}}
\hfil
\caption{(a) Papers evaluating four environmental factors (legend and different line colors) with intervals for year of publication on the x-axis and the total number of papers for the range of each year interval on the y-axis. (b) Environmental factors are shown in the legend with different bar colors. The topic areas for papers are on the x-axis and the total number of papers shown on the y-axis.}
\label{fig:Factors}
\vspace{-1em}
\end{figure*}

\section{Environmental Factors}\label{EnvironmentalFactorsSection}
In this section, we discuss some common environmental factors, such as sunlight and surrounding color difference between indoor and outdoor, that may affect the results of HMD measurements, and their prevalence in the surveyed literature, illustrated in Figures \ref{fig:FactorsYear} and \ref{fig:FactorsSummary}.

\subsection{Indoor Environment vs. Outdoor Environment}
One of the most frequently mentioned environmental factors is the difference between indoor and outdoor environments. 
For example, \cite{Erickson20} mentions that most OST-HMDs add AR light so they do not have to compete against real-world lighting. However, when the OST-HMDs are exposed to direct sunlight in outdoor environments, they lose a significant amount of contrast. The authors recommend that beyond reducing environment light, future OST-HMD works need to consider simulating and making adjustments to the intensive environment lighting in the outdoors.
In a comparison of egocentric distance judgments with two different HMDs using indoor and outdoor virtual environments, \cite{Creem15} found greater underestimation in outdoor than indoor environments (see also \cite{Buck18}). 
In addition, \cite{Grechkin10} describes how the typical settings of indoor and outdoor environments should be considered; for example, hallways are typical for indoor settings, while lawns are typical for outdoor. 
In \cite{Liu2021}, the experimenters tested their HMD-based system in both indoor and outdoor environments to achieve complete 3-D eye-tracking accuracy.
Lastly, \cite{Lee22} used a VR device to simulate the MR world and compared the usability of ten different navigation instruction designs.

While the differences between indoor vs. outdoor environments are frequently discussed in the literature, Figure \ref{fig:FactorsYear}, shows that this factor in the first two time intervals is actually assessed in measurement studies relatively infrequently. Indoor experimental settings are often more convenient and afford more experimental control, especially over variables like luminance. HMD measurement experiments that compare outdoor and indoor environments faced significant challenges in the past. However, since the application of HMDs to the outdoors is a critical challenge, this factor became more prevalent after 2010, and we expect the same trend to continue in future studies.

\subsection{Virtual Environment vs. Virtual/Real Environment}\label{VEvsVRE} 
Another important factor for measurement studies is the difference between two virtual environments or between virtual and real environments. Under the transition of different environments, the virtual objects rendered in front of the user may be displayed differently through HMDs. With different rendering effects, experimenters can do the same measurement across different environments to detect different results. 

In \cite{Kijima97}, the authors developed a Projective Head Mounted Display (PHMD) to build a compound environment of the workstation and surrounding virtual world.
In addition, \cite{Willemsen2004, Willemsen09} designed an experiment under real and virtual room conditions. The result indicated that the mock HMD in the real-world caused a reliable effect of underestimation.
Then, \cite{Hiramoto18} proposed an idea of how to compensate for the difference in space perception between the real-world and the VR environment by asking users to reach the farthest position in each environment. 

Figure \ref{fig:FactorsYear} shows that a large proportion of included papers compare either virtual vs. virtual environment or virtual vs. real environment. 
One reason to compare two virtual environments is to ensure that results (e.g. differences between two different HMDs) generalize beyond a single virtual environment.  
When virtual vs. real environments are compared, the main goal is usually to observe that the participants produce different responses and feedback to the same type of display in these two environments. This experiment allows the researchers to understand that the human response to the real environment may not be exactly the same as the response of the virtual environment, and more attention and understanding are needed when designing HMD-related work.

\subsection{Color Difference}
Similarly, recent papers also consider the difference, or lack thereof, for the color of virtual objects and color of the environment. Rendering virtual objects with colors close to the color of environment may make it difficult for participants to see the target objects created by HMDs. This could also impair judgements of distance.

The authors of \cite{Itoh2015} noted frequent errors in the AR application of OST-HMDs between the input and output colors on display, and applied a complete color reproduction method to solve this problem. 
In \cite{Alamri2010}, the experimenters used different colors to indicate different target positions for the participants.
In addition, \cite{Harding2018} introduced an extended color discrimination model to address the issue of the combination of the color displayed by the HMD and the color of the external environment leading to a false perception.
By using optical and video see-through HMDs as evaluation devices in the MR world, \cite{Shin2020} examined the effect of color parameters on users’ perceptions when using different types of MR devices, showing that users often overestimated color in MR environments.
Finally, in recent years, high-fidelity color in MR environments has been an essential task for every OST-HMD. \cite{Zhang2021} proposed a rendering method to enhance the color contrast between virtual objects and the background of the real environment, which can raise the fidelity degree of color in MR environments and provide a better visual experience.

Figure \ref{fig:FactorsYear} shows that older papers exploring the effects of color differences are rare, with more appearing recently. This may be explained by the relatively low resolution of older-version HMDs. The increased resolution of new versions of HMDs (for example, the display resolution of HoloLens 2 is 2048x1080 per eye, and the display resolution of Oculus Quest is 1440×1600 per eye) may mean that users are more sensitive to the colors displayed in HMDs.
In future measurement-related research in AR/MR, the possible impact of virtual color with the real-world environment is likely to be increasingly important.

\subsection{Luminance Difference}
The luminance of the environment, both alone and in combination with color, is also an important factor for the perception of virtual objects. Here, luminance indicates the intensity of light in the environment. For example, a dark red background will make bright green easier to notice, or excessively high luminance may prevent users from seeing the target object clearly in a dark environment.

In \cite{Li2018}, the authors designed a series of experiments to examine the relation between luminance and distance judgments in HMDs, such as finding a certain threshold of the luminance that affects the result.
In addition to luminance affecting distance judgment, \cite{Ha2020} proposed that significant increases in luminance while viewing HMDs can cause visual discomfort. This paper evaluated the luminance change and maximum luminance that will cause discomfort to the users.
Recent research has focused specifically on environment luminance affecting OST-HMDs, which use additive light models, meaning they lose a significant amount of contrast in high luminance environments (e.g., outdoor on a sunny day with no clouds). In \cite{Erickson20}, the authors designed a measurement to evaluate perceptual challenges with OST-HMDs and made some suggestions for future improvements.

As shown in Figure \ref{fig:FactorsYear}, studies have assessed luminance differences as a factor throughout the time periods surveyed, with what appears to be a small increase over time. Common effects of different luminances are changing visibility of virtual objects for users, and high-luminance environments causing user discomfort. 
As mentioned above, indoor and outdoor luminances are often substantially different. Some related papers used indoor high luminance to simulate outdoor luminance while controlling for other differences between indoor and outdoor environments. Others have recently tried to make HMDs usable in both indoor and outdoor environments. 

We also summarize the number of papers for each environmental factor over time and the specific topic areas for environmental factor in Figure \ref{fig:FactorsSummary}. 
This figure again shows the relatively high proportion of papers with the category of Virtual/Real, across multiple topic areas. 
By topic area, environmental interaction and questionnaires are frequently covered in papers assessing all four environmental factors. Whereas the topic area of comfort assessment only has three papers, two of which focus on the luminance (magenta bar) conditions that cause participants to experience discomfort while wearing HMDs. 
Similarly, eye tracking only has a single paper for each of three environmental factors (color, luminance, and in/out) since attention-drawing factors used in these papers are often the changes in color or luminance, these are common environmental factors when designing the measurement of eye tracking.

\section{Existing experiment designs among measurements}\label{ExistingExperiments}
HMD-related research using AR, VR, and MR uses a variety of experimental designs, depending on the research question, including measurements to test the machine's performance or to assess human response to specific aspects of the virtual environment. 
In this section, we summarize the experimental design of the included papers according to three frequent experimental factors: participants, data collection, and mechanism.

\subsection{Participants}\label{ParticipantsSub}
In this subsection, we examine the relationships among the number of participants recruited for measurement studies, the composition of participant samples, and the category of the studies. For example, many studies are conducted at universities \cite{Dey2018}, meaning that participants are largely students who are less than 30 years old. We explore the distribution of participant ages among different categories of measurement papers.
Similarly, men generally outnumber women among students in computer science and engineering. There are potentially gender differences in the results of the experiments, so we also compare the male and female ratios of participant samples in the surveyed papers. 
Previous HMD experience may also potentially affect results; we organized papers by whether they assessed HMD experience in their participants and if so, the distribution of experience. 

\begin{figure*}[h]
\centering
\includegraphics[width=0.7\textwidth, keepaspectratio=true]{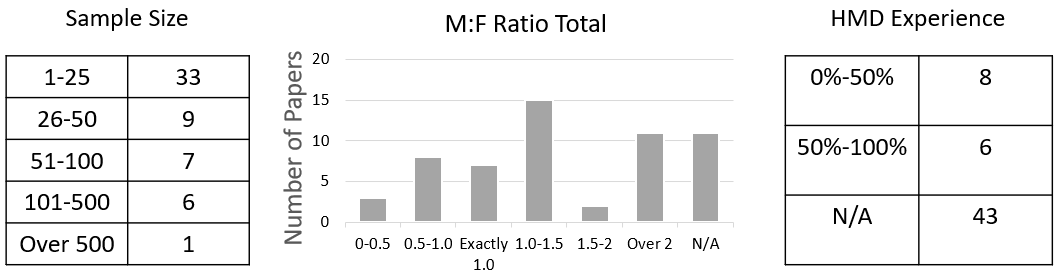}
\caption{Sample size, Male-to-Female ratio, and HMD experience in related environmental measurement papers}
\label{fig:ParticipantsInfo}
\vspace{-1em}
\end{figure*}

\begin{figure*}[b]
\centering
\includegraphics[width=0.7\textwidth, keepaspectratio=true]{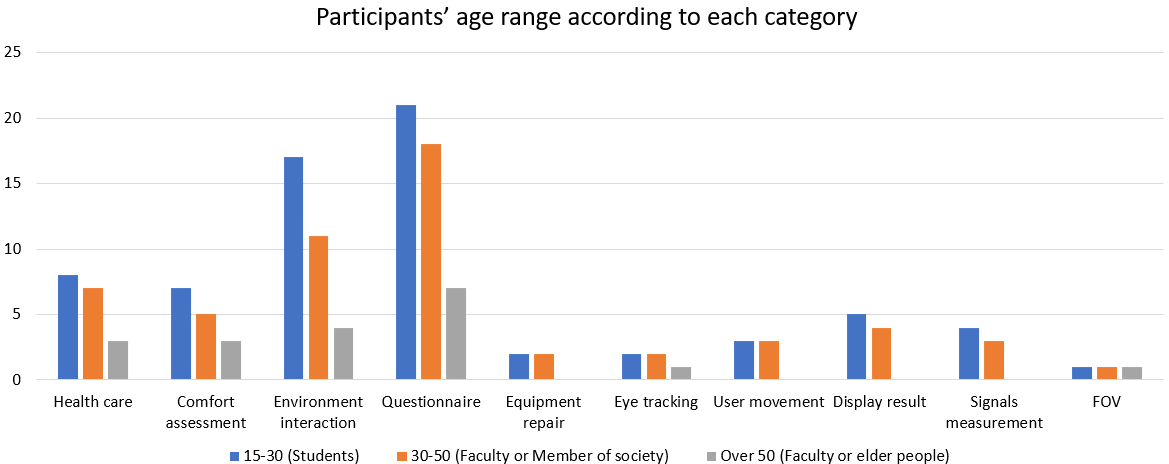}
\caption{Participant age ranges for each topic area for related environmental measurement papers}
\label{fig:ParticipantsAge}
\vspace{-1em}
\end{figure*}

The left panel of Figure \ref{fig:ParticipantsInfo} shows the ``sample size'' or total number of participants recruited in each paper. Note in the majority of papers the sample size tends to be limited, between \emph{N} = 1 to 25. 
Many factors may affect the number of participants, such as: aims of the study (e.g., qualitative usability vs. quantitative performance), study design (e.g., whether or not there are repeated measures), study duration and  resources needed to run the study, and whether participants with particular expertise is required. Nevertheless, in general, larger sample sizes are desirable for quantitative research using inferential statistics.

The Male-to-Female Ratio Total chart in the middle panel of Figure \ref{fig:ParticipantsInfo} does not support the above assumption that participant samples in most experiments are overwhelmingly male. Although there may be specific areas of research such as male athlete sports-related \cite{Cummiskey17} that will include only male participants, or some papers that do not describe the ratio of men to women in detail, most of the measurement-related papers attempted to roughly balance the ratio of men to women. To ensure that experimental results generalize across genders, it is important to recruit both male and female participants. However, it can be challenging to achieve a male-to-female ratio of exactly 1. We consider an M:F ratio between 0.5 to 1.5 to be roughly balanced; Figure \ref{fig:ParticipantsInfo} shows that most papers fall within this guideline.

Prior experience with HMDs is also an important participant variable. As shown in the right panel of Figure \ref{fig:ParticipantsInfo}, most papers did not report participant HMD experience (marked as "N/A"). The small number of papers (14 out of 57) that did report participants' prior experience with HMDs were roughly balanced as to whether the majority of their sample had experience with HMDs. Depending on the category of the measurement experiment, the HMD experience of the participant who needs to be recruited will be different since the participants' experience with HMDs may have different progress and outcomes for the experiment. The most common case is in the category of health care, where recruiting experienced participants enables experiments to be performed more quickly and accurately because no additional training is required for them.

As \cite{Dey2018} proposed, since most academic papers are written in universities the age distribution of most participants tends to skew younger, corresponding to the typical age of university students. Figure \ref{fig:ParticipantsAge} depicts the age range of participants in papers from different topic areas. Most papers provided an age range of participants, but only a small number of papers reported more detail such as the mean age or the source of the participant sample (e.g., university students). In support of \cite{Dey2018}, we see that the age range between 15-30 in many topic areas is relatively high, likely because recruiting students in universities is convenient and inexpensive when the experiment requires many participants. Also, experiments with novel HMD technology may be more likely to attract students' attention and make them more willing to participate in the experiment. We do not see evidence in Figure \ref{fig:ParticipantsAge} that the age distribution of participants is related to the paper category; most of the researchers recruited 15- to 30-year-old students as participants for each category.
However, participant age can often have an impact on experimental results. Much like balancing Male-to-Female ratios, it is important to assess a range of participant ages in order to be able to generalize results. Furthermore, there may be research applications where older users who are relatively unfamiliar with HMDs are the relevant population. Therefore, we recommend that researchers report detailed information about the ages of their participants, and also consider recruiting beyond the typically convenient population of university students. 

\begin{figure*}[t]
\centering
\includegraphics[width=0.6\textwidth, keepaspectratio=true]{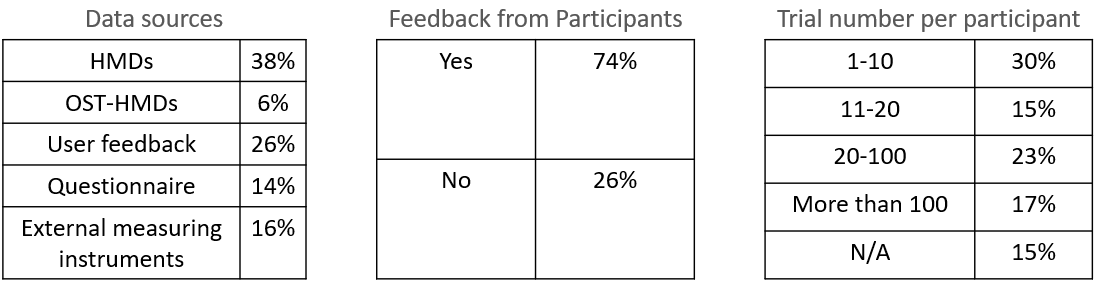}
\caption{Data sources of measurement experiments (Left), whether or not participants gave feedback to experimenters (Center), and experiment trial number for each recruited participant in related measurements (Right) (*OST-HMDs = Optimal see-through Head-Mounted Devices)}
\label{fig:DataSourceTrialNumAndFeedback}
\vspace{-1em}
\end{figure*}

\subsection{Data Collection}\label{DataCollection}
The method of data collection is a critical aspect of any measurement experiment, and there are a number of potential data sources for studies with HMDs, including sensor data (HMDs or external measuring instruments), user responses, and questionnaires. We collected the data acquisition methods in measurement papers related to HMDs using AR and MR and organized them in this section, and then summarized this  information in Figure \ref{fig:DataSourceTrialNumAndFeedback}.

\begin{figure*}[b]
\centering
\subfloat[]{\includegraphics[width=0.4\textwidth, keepaspectratio=true]{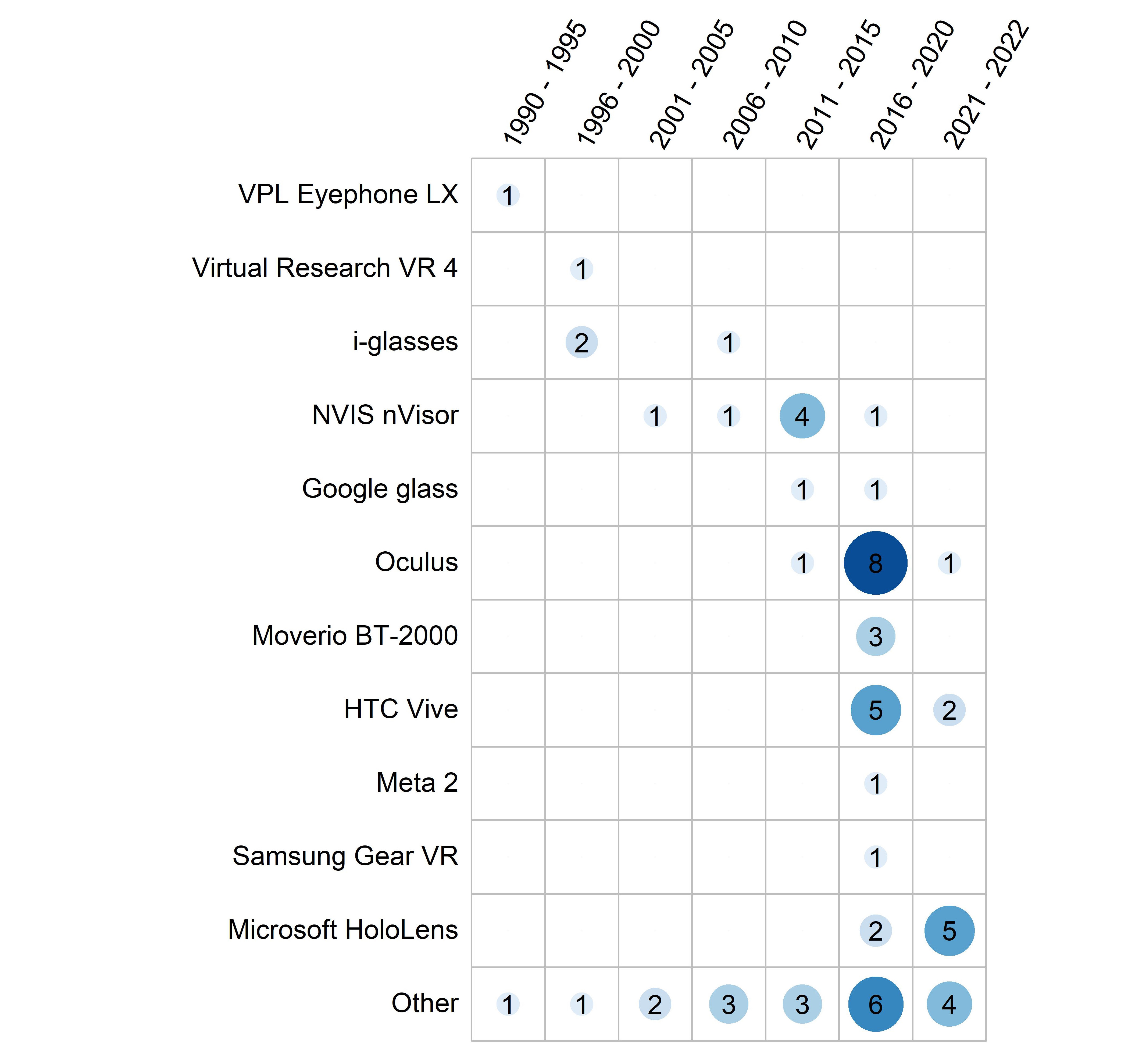}%
\label{fig:HMDsYears}}
\hfil
\subfloat[]{\includegraphics[width=.4\textwidth, keepaspectratio=true]{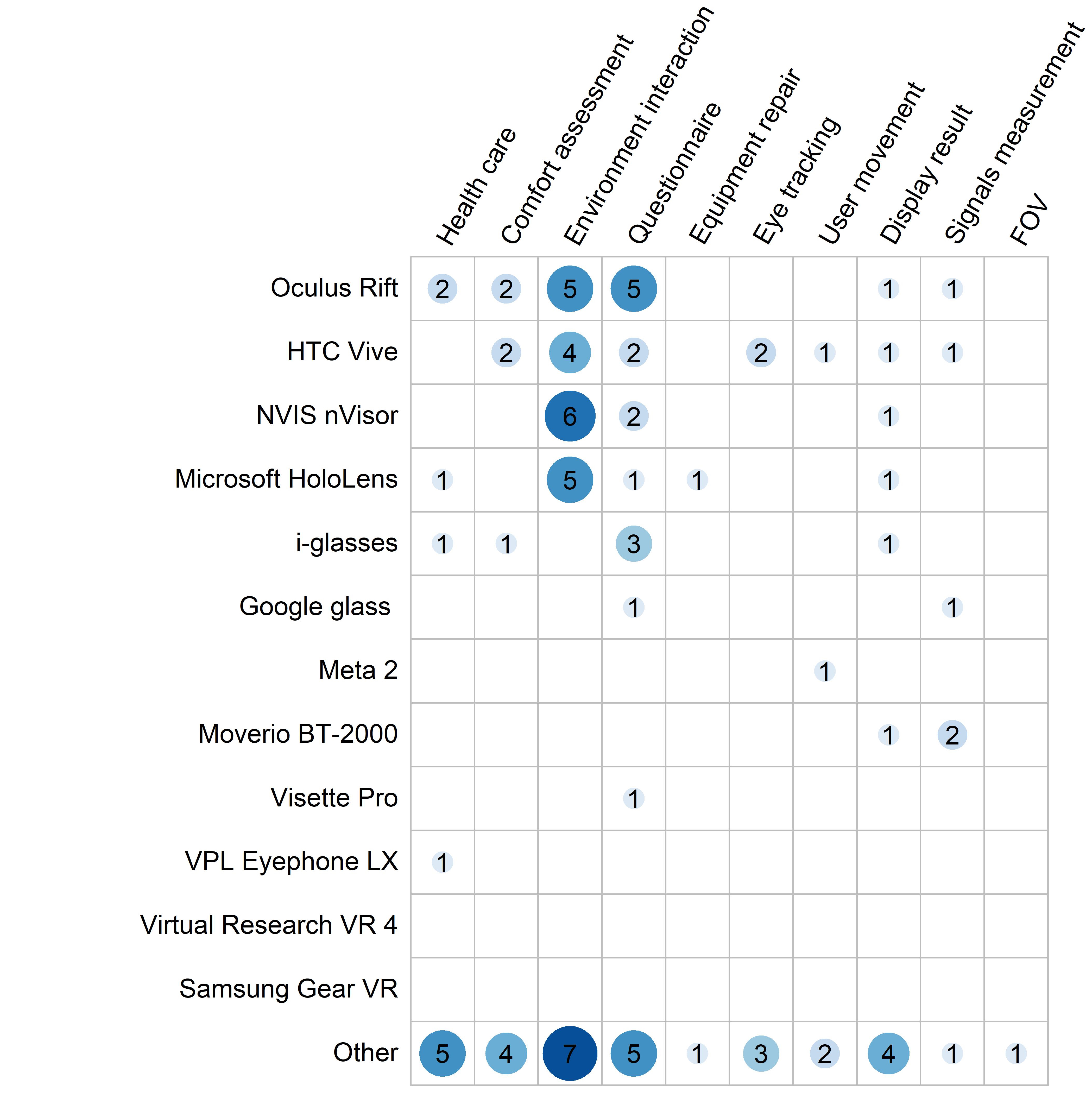}%
\label{fig:HMDsCategory}}
\hfil
\caption{(a) Head-Mounted Devices used in related measurements according to different years. (b) Head-Mounted Devices used in related measurements according to different categories.}
\label{fig:HMDUsed}
\vspace{-1em}
\end{figure*}

Although HMDs are the most common data source in the surveyed papers, since measurements often require participants input, many papers also use user feedback as a data source. User feedback is usually divided into verbal feedback during the experiment, such as verbally telling the experimenter the distance to a target, or written feedback, such as completing questionnaires after the test. Since questionnaire was commonly used in measurement-related papers, we created a separate questionnaire category in Figure \ref{fig:DataSourceTrialNumAndFeedback} (Left). Other standard data sources include external sensors that record data in HMDs or computer software that connects to HMDs to collect data; we collectively call these "External measuring instruments." Furthermore, our survey indicated that many recent measurement papers used OST-HMDs as their main measurement data sources. Therefore, we created OST-HMDs as an additional category in Figure \ref{fig:DataSourceTrialNumAndFeedback} (Left). 

In addition, \ref{fig:DataSourceTrialNumAndFeedback} (Center) shows the majority of papers directly assessed feedback from participants, which can be used to improve the experiment process, time length, or other environmental factors based on the feedback.
Then, about half of the papers had 11 trials or more per participant, shown in Figure \ref{fig:DataSourceTrialNumAndFeedback} (right). Note that 15\% of papers did not report the number of trials for each participant, marked as "N/A." Only 17\% of papers had $>$100 repeated trials per participant, this may be because of the possibility of fatigue and/or discomfort with extended use of an HMD.

\subsection{Mechanism}\label{MechanismSubSection}
As of early 2022, there are many different types of HMDs on the market, including AR, VR, and MR HMDs. Different HMDs use different computer platforms and sensors, and experimenters need to design corresponding measurement experiments according to the capabilities of the device. In this section, we summarize the assessment of different HMDs across time and various topic areas.
Since the number of HMDs for AR and MR is relatively small, we also include papers using VR HMDs in this section.

Figure \ref{fig:HMDsYears} shows that the most widely used HMDs are different depending on the year. In earlier papers, there are many different types of HMDs used, even HMDs built by the researchers themselves. Since about 2004, authors of measurement-related papers began to use one of the most common HMDs during that time, which was a different version of NVIS nVisor \cite{NVISWeb}. After about 2014, other modern and well-known HMDs, such as Google glass \cite{GoogleGlassWeb}, Oculus \cite{RiftWeb}, and HTC Vive \cite{HTCViveWeb}, began to appear and were widely used by researchers. In addition to general HMDs, OST-HMDs have been frequently used in recent years; such devices include Microsoft HoloLens 1st and 2nd generation \cite{hololensWeb}, which is clearly shown in the rightmost columns of Figure \ref{fig:HMDsYears}. Researchers' widespread use of OST-HMDs means that MR has begun to flourish, and more information close to reality can be obtained. In addition, although not mentioned in detail in this figure, a new version of Oculus, Oculus Quest, has been widely used in research in recent years. Oculus Quest is wireless and thus will be more helpful to future research. Since a large number of papers use custom HMDs or rarely used HMDs, we classify them in the "other" category. For example, in \cite{Schneider2005}, the authors used swimming goggles, video cameras, transparent hot mirrors, and eye trackers to build a custom HMD. There are many papers focusing on technological advances in HMDs, and the use of custom HMDs enables researchers to break away from existing HMDs to achieve their goals. Therefore, the proportion of the category "Other" in Figure \ref{fig:HMDsYears} and Figure \ref{fig:HMDsCategory} is relatively high. 

\begin{figure*}[t]
\centering
\includegraphics[width=0.7\textwidth, keepaspectratio=true]{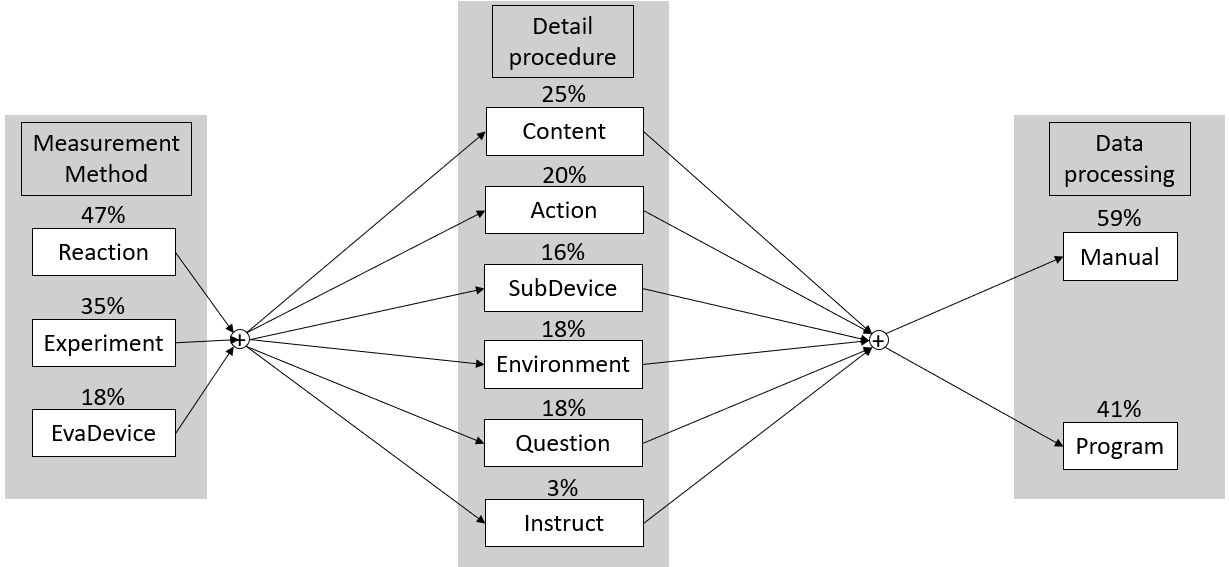}
\caption{Experimental procedures used in related measurement papers and proportion of related papers using these procedures. (Reaction: Participants’ reaction during test, Experiment: Participants’ user experiment after test, EvaDevice: Device evaluation, Content: Content change, Action: Participants' action, SubDevice: Device substitution, Environment: Environment change, Question: Questionnaire, Instruct: Experimenter instruction, Manual: Manual observation, Program: Computer program calculation)}
\label{fig:ExpProcedure}
\vspace{-1em}
\end{figure*}

Figure \ref{fig:HMDsCategory} shows the HMDs used across different measurement categories. There is no clear relationship between category and device type. There are many experiments of the same type that can be used in similar HMDs; for example, \cite{Buck18} in 2018 used several different HMDs in the topic of distance underestimation, including Oculus Rift, Nvis HMDs, and HTC Vive. We see again a large number of devices falling in the "other" category in this figure. As described above, many of these are custom HMDs, which were often designed only for a specific research field. For example, in \cite{Culham04} in 2004, the target research area was health care, so the HMDs used needed to match the medical needs to select the appropriate machine.

\section{Experimental Procedure}\label{ExperimentalProcedureSection}
In this section, we summarize frequently used experimental procedures across the literature (depicted in Figure \ref{fig:ExpProcedure}), including:\\
\begin{itemize}
\item \textbf{Measurement method}: classifies papers based on whether there are human participants and whether they provide data during or after the test.
\item \textbf{Detail procedure}: categorizes the design details of the experiment into: Content change, Participants' action, Device substitution, Environment change, and Questionnaire. 
\item \textbf{Data processing method}: describes how the experimenter processes the data after collecting the data, manually or with 
Computer program calculation.
\end{itemize}

In Figure \ref{fig:ExpProcedure}, we first divide the measurement method mentioned into two main categories: \textbf{device evaluation} (EvaDevice) (\cite{Mine1993, Javier14}) and \textbf{participants' reaction}. We further divide the category of participants' reaction into \textbf{Participants' reaction during test} (Reaction) (\cite{Willemsen2004, Batmaz2019}) and \textbf{Participants' user experiment after test} (Experiment) (\cite{Zhang2012, Hirzle2021}). Most measurement-related papers recruit participants to obtain user feedback. Sometimes participants' reactions are collected during the experiment, and sometimes participant feedback is recorded after completing the experiment. Some measurement papers do not recruit participants but only conduct tests to compare differences between different HMDs under certain factors, so we call this category \textbf{device evaluation}. Dividing into these three categories allows researchers to quickly understand how most measurement-related papers design their experimental procedures and refer to these related works' methods in their own designs. 

In addition, the center of Figure \ref{fig:ExpProcedure} shows the six main categories of detailed procedure we encountered in the surveyed literature. These categories include: content change (Content; \cite{Naceri2009, Leyrer2015, Arpaia2020, Angrisani2020}), participants' action (Action; \cite{Koulieris2017, Kim2012}), device substitution (SubDevice; \cite{Messing2005, Adams21, Heinrich22}), environment change (Environment; \cite{Nakao14, Harding2018}), questionnaire (Question; \cite{Nichols00, Singla2017}), and experimenter instruction (Instruct; \cite{Grechkin10, Wittich2018}). 
First, \textbf{Content change} is to change the settings of items in the test environment, such as the color, luminance, or size of virtual/real objects. This procedure is the most common (25\%) in measurement-related papers perhaps because modifying the shape or attribute of an object is an intuitive and accessible change.
Second, \textbf{participants' action} refers to participants taking corresponding actions under the experimenter's preset settings. For example, the participant moves the target object from one location to another under the preset settings of the experiment. This procedure is also very common (20\%) among measurement-related papers. 
The \textbf{device substitution} category refers to experiments that  test and compare the results produced by different HMDs in the same situation. These results can inform the selection of appropriate HMDs to be used in future measurement-related experiments or implementations.
The concept of \textbf{environment change} includes papers that manipulate the test environment where the assessment occurs during the experiment. As described in the Virtual Environment vs. Virtual/Real Environment subsection (\ref{VEvsVRE}), the impact of the environment on participants and HMDs can be substantial. 
Next, one of the most commonly used methods for obtaining user feedback, \textbf{questionnaire}, is classified as the fifth procedures category. Questionnaires are typically administered after the experiment to obtain user feedback, especially in the health care and comfort assessment categories. However, there are also a small number of experiments that use questionnaires to obtain some user information and user expectations before the experiment starts. Questionnaire information can be extremely useful for assessing individual differences in participants as well as usability and preferences. 
Finally, a rare (3\%) but important procedure used in HMD measurement-related papers is \textbf{experimenter instruction}. Similar to the Action category, the experimenters instruct the participant to do specific actions during the experiment, but in this procedure the experimenter provides different instructions according to the different reactions of each participant. This method is reasonably practical in measurement, and changing the direction of the experiment according to different situations can allow for different insights than can otherwise be obtained. 

In the right side of Figure \ref{fig:ExpProcedure}, we divide the data processing method into two categories: manual observation (Manual) (\cite{Peli1998, Shin2020}) and computer program calculation (Program) (\cite{Moghimi2016, Do20}). These two methods represent how the experimenter processes the data obtained via the measurement method and detailed procedure mentioned above. 
\textbf{Manual observation} means that after obtaining the feedback data from the participant, the experimenter manually sorts and analyzes the data. A common method is to manually code process subjective feedback data from questionnaires. The advantage of this method is that in collation and analysis, the experimenter may uncover unanticipated patterns in participant responses. However, the disadvantage is that it takes more time to process the data.
\textbf{Computer program calculation} refers to the use of computer programs to organize and output the results needed by the experimenter after the data is obtained. The advantage is that it can quickly organize all the information and produce easy-to-read results, such as graphs or tables. The disadvantage is that it is easy to overlook some subtle information that is not the main experiment goal, which may lead to incomplete or misleading results.
We found that papers that used the manual observation method are more common (59\%) than papers that used computer program calculation (41\%), perhaps because manual observation is more straightforward and can be implemented relatively quickly.

\section{Challenges, Limitations, and Discussion}\label{ChallengesSection}
Researchers will face a variety of challenges in the future, such as insufficient machine technology or difficulty simulating the real environment fully in the experimental virtual environment. Note that achieving high realism is not necessarily beneficial for effective performance in simulation-based training for aviation and health care, despite strong user preferences for high visual fidelity \cite{drews2013simulation}. Nevertheless, as time goes by, some of the challenges encountered in previous papers are likely to be solved by researchers and documented in future papers. Therefore, in this section, we introduce several common challenges and possible solutions in measurement experiments related to HMDs.

Several of the surveyed measurement papers discussed the challenges and limitations they encountered.
In a paper on environmental changes, \cite{Nakao14} mentioned that they should consider whether the experiment could proceed as smoothly when switching environments in their future work, which shows that the aforementioned environmental factors of virtual environment vs. virtual/real environment are essential and challenging. Furthermore, they suggested that using more accurate medical equipment on top of a questionnaire would be more accurate; this highlights the benefit of using multiple data sources.
In \cite{Itoh2015}, the authors discussed the limitations in the method and hardware they faced for their OST-HMD color calibrations method, including both eye model issues and HMD model issues. For the eye model issues, due to technical limitations there was and is no technology to fully understand how the human retina absorbs sunlight and allows the human brain to perceive colors. Therefore, even OST-HMD cannot fully simulate the response of the human retina. The HMD issue refers to the limited display color range of the OST-HMD and the interference with the environmental light that can cause display distortion.

Another limitation mentioned by \cite{Erickson20} is that the color contrast will decrease when using OST-HMD in high-brightness indoors or outdoors. These and other authors \cite{Do20} also describe the limitations due to the outbreak of COVID-19, such as difficulties recruiting large enough participant samples for sufficient experimental data. If future works can develop methods to perform HMD experiments remotely, this will be extremely beneficial if a similar situation occurs again. 
Perhaps related to technical limitations of current HMDs, such as weight and FOV, perceived distances tend to have greater underestimation outdoors than indoors. For example, \cite{Creem15} stated that the research on distance underestimation in outdoor environments had not been fully explored since outdoor environments have numerous variables (e.g., overall scale, realism, and geometrical complexity). In mixed reality, the misperception of space may also depend on the distances. Using mixed reality, \cite{gagnon2020far} evaluated far distance perception. They found both overestimation (at 25 to 200 meters) and underestimation (at 300 to 500 meters) of distances. To the best of our knowledge, use of HMDs for far distances, whether indoor or outdoors, is a research gap. It is also a technological limitation of current devices (e.g., the optimal distances for projecting AR content with a Hololens 2 is 1.25 to 5 meters from the user \cite{hololensWeb}).      

The authors in \cite{Li2018} mention the limited FOV users experience when wearing an HMD, making it difficult to reproduce the same experience of the real world in HMD. If a primary goal of HMD development is to increase user comfort with the devices, an HMD with all-covered FOV will be a critical goal.
Several authors \cite{Shu19, Sharples2008, Moss11} mention the problem that while HMDs may provide a greater sense of presence to users but can cause simulator sickness. A recent systematic literature review of more than 300 papers \cite{Teresa21} found that a widely used measure of simulator sickness did not fully capture other comfort factors (e.g., eye strain and physical comfort). Therefore, a range of side effects should also be considered in the experiments' design to avoid fatigue, dizziness, and other discomforts that may interfere with experimental results. Many researchers are now trying to solve these problems. 

As summarized above, many common challenges and limitations are due to the current technical capabilities of HMDs, with the recent majority related to color (display) and luminance (Outdoor environment) in environmental factors. Therefore, solving existing challenges can directly affect the development of future HMDs and bring better experiences to users.

\section{Conclusion}\label{ConclusionSection}
\subsection{Summary}
In this paper, we have categorized a total of 87 papers using HMDs in AR, MR, and VR. The categories included environmental factors and user vs. device measurements. We identified several potential research gaps. Comparison of indoor vs. outdoor was the environmental factors assessed by the fewest papers. This may be largely due to technical limitations with the performance of current HMDs in outdoor environments, but the number continues to increase after 2010. In addition, only one paper evaluated user movement. This indicates almost all of the papers had participants in a stationary standing or sitting position. However, real-world use of AR/MR for tasks such as maintenance, some health care applications (e.g., surgery), or spatial navigation will typically require walking and/or changing body positions (e.g., leaning to look from different perspectives). In the device measurement category, no papers proposed standardized methods for evaluating and comparing technical aspects of HMDs. 

In terms of gaps for user experiments, we also found these papers tended to have limited sample sizes. The majority of papers had 25 or fewer participants, although many papers had repeated measures (multiple trials per participant). Surprisingly, there was a clear lack of a gap for the ratio of male to female participants which were often similar. However, most papers did not report whether they assessed prior HMD experience. If HMD use continues to grow as predicted over time, assessing experience with HMDs is likely to become increasingly useful. Additionally, most studies tended to rely on a younger, student population for participant samples, which may not be ideal for all applications or for generalizability of results.

In terms of hardware, the most frequently used HMDs were the NVIS nVisor, Oculus, Google Glass, and Microsoft HoloLens. There were many "other" devices and seven different HMDs were only used in a single paper. 
While we categorized HMD hardware, a limitation of the current paper is that we cannot fully categorize the software used by paper due to large software diversity (e.g., custom software versus commercial off the shelf software such as Unity or Unreal Engine versus open source such as ARToolKit). But we still list the software mainly used by mentioned HMD in the Appendix \ref{ThirdAppendix}. 

\subsection{Future Directions}
In the future, researchers can find the reference materials they need based on this paper and use our integrated figures to find trends in these research topics in recent years. We hope this paper can be an informative overview to help further develop categories and experiments with environmental-related measurements. Furthermore, this categorization may also be useful for other different types of AR/MR devices, such as mobile AR/MR devices.

\appendices

\section{References List For Each Environmental Factor}\label{FirstAppendix}

\begin{center}
\small
\begin{tabular}{ |c|c|c| } 
 \hline
 \textbf{In/Out} & 1990-1999 & (total 0) \\ 
 \hline
 (Total 11) & 2000-2009 & \cite{Renkewitz2008}, \cite{Grechkin10} (total 2) \\ 
 \hline
  & 2010-2019 & \cite{Buck18}, \cite{Nakao14}, \cite{DeFelice2014}, \cite{Rangelova2018}, \cite{Erickson20}, \cite{Weber2021}, \\ 
  & & \cite{Creem15} (total 7) \\ 
 \hline
  & 2020-2022 & \cite{Liu2021}, \cite{Lee22} (total 2) \\ 
 \hline
 \textbf{Virtual/Real} & 1990-1999 & \cite{MonWilliams1993}, \cite{Peli1998}, \cite{Kijima97} (total 3) \\ 
 \hline
 (Total 33) & 2000-2009 & \cite{Santos09}, \cite{Patrick2000}, \cite{Mohler2007}, \cite{Willemsen2004}, \cite{Willemsen09}, \cite{Naceri2009} \\
 & & (total 6) \\ 
 \hline
  & 2010-2019 & \cite{Buck18}, \cite{Profita16}, \cite{Nakao14}, \cite{DeFelice2014}, \cite{Rangelova2018}, \cite{Robert2016}, \\
  & & \cite{Chan2019}, \cite{Shu19}, \cite{Peer2017}, \cite{Creem15}, \cite{Hiramoto18}, \cite{Itoh2015}, \\
  & & \cite{Zhang2012}, \cite{Leyrer2015}, \cite{Koulieris2017}, \cite{Kim2012}, \cite{Singla2017}, \\
  & & \cite{Moghimi2016} (total 18) \\ 
 \hline
  & 2020-2022 & \cite{Lin22}, \cite{Khan21}, \cite{Ariansyah2022}, \cite{Weber2021}, \cite{Adams21}, \cite{Heinrich22} \\ 
  & & (total 6) \\ 
 \hline
 \textbf{Color} & 1990-1999 & (total 0) \\ 
 \hline
 (Total 13) & 2000-2009 & \cite{Renkewitz2008}, \cite{Naceri2009} (total 2) \\ 
 \hline
  & 2010-2019 & \cite{Wittich2018}, \cite{Wille2013}, \cite{Alamri2010}, \cite{Itoh2015}, \cite{Harding2018}, \cite{Moghimi2016} \\
  & & (total 6) \\ 
 \hline
  & 2020-2022 & \cite{Shin2020}, \cite{Do20}, \cite{Zhang2021}, \cite{Imaoka2020}, \cite{Adams21} \\
  & & (total 5) \\ 
 \hline
 \textbf{Luminance} & 1990-1999 & \cite{MonWilliams1993} (total 1) \\ 
 \hline
 (Total 12) & 2000-2009 & \cite{Renkewitz2008}, \cite{Birkfellner2002} (total 2) \\ 
 \hline
  & 2010-2019 & \cite{Cognolato2018}, \cite{Wang2019}, \cite{Li2018}, \cite{Zhang2012}, \cite{Leyrer2015} \\ 
  & & (total 5) \\ 
 \hline
  & 2020-2022 & \cite{Shin2020}, \cite{Kahl21}, \cite{Ha2020}, \cite{Erickson20} (total 4) \\ 
 \hline
\end{tabular}
\end{center}

\section{References List For Each Category}\label{SecondAppendix}

\begin{center}
\small
\begin{tabular}{ |c|c| } 
 \hline
 \textbf{Health care} & \cite{Wittich2018}, \cite{Culham04}, \cite{Cummiskey17}, \cite{Birkfellner2002}, \cite{Chan2019}, \cite{MonWilliams1993}, \cite{Peli1998}, \\ 
  & \cite{Alamri2010}, \cite{Koulieris2017},  \cite{Heinrich22}, \cite{Nichols00} (total 11) \\
 \hline
 \textbf{Comfort assessment} & \cite{Arefin22}, \cite{Rangelova2018}, \cite{Robert2016}, \cite{Ha2020}, \cite{Wille2013},  \cite{Hirzle2021}, \\ 
  & \cite{Koulieris2017},  \cite{Singla2017} (total 8) \\ 
 \hline
 \textbf{Environment interaction} & \cite{Buck18}, \cite{Lin22}, \cite{Nakao14}, \cite{Kahl21}, \cite{Swan06}, \cite{Khan21}, \\ 
  & \cite{Wille2013}, \cite{Peer2017}, \cite{Erickson20}, \cite{Mohler2007}, \cite{Creem15}, \cite{Lee22}, \\
   & \cite{Willemsen2004}, \cite{Hiramoto18}, \cite{Li2018}, \cite{Zhang2012}, \cite{Naceri2009}, \cite{Leyrer2015}, \\
    & \cite{Kim2012}, \cite{Messing2005}, \cite{Adams21}, \cite{Moghimi2016} (total 22) \\ 
 \hline
 \textbf{Questionnaire} & \cite{Santos09}, \cite{Profita16}, \cite{Lin22}, \cite{Arefin22}, \cite{Nakao14}, \cite{Rangelova2018}, \\
  & \cite{Patrick2000}, \cite{Shin2020}, \cite{Shu19}, \cite{Kahl21}, \cite{Khan21}, \cite{Peli1998}, \\ 
   & \cite{Ha2020}, \cite{Peer2017}, \cite{Wang2019}, \cite{Batmaz2019}, \cite{Lee22}, \cite{Zhang2012}, \\
    & \cite{Hirzle2021}, \cite{Leyrer2015}, \cite{Kim2012}, \cite{Adams21}, \cite{Nichols00}, \\ 
     & \cite{Singla2017} (total 24) \\ 
 \hline
 \textbf{Equipment repair} & \cite{Ariansyah2022} (total 1) \\ 
 \hline
 \textbf{Eye tracking} & \cite{Cognolato2018}, \cite{Wang2019}, \cite{Imaoka2020}, \cite{Lidegaard14}, \cite{Liu2021}, \cite{Schneider2005} \\ 
 & (total 6) \\ 
 \hline
 \textbf{User movement} & \cite{Mohler2007}, \cite{Batmaz2019} (total 2) \\ 
 \hline
 \textbf{Display result} & \cite{Santos09}, \cite{Patrick2000}, \cite{Shin2020}, \cite{Do20}, \cite{Zhang2021}, \cite{Shu19}, \\
 & \cite{Batmaz2019}, \cite{Mine1993}, \cite{Watson1997}, \cite{Kim2012}, \cite{Singla2017} \\ 
 & (total 11) \\ 
 \hline
 \textbf{Signals measurement} & \cite{Alamri2010}, \cite{Weber2021}, \cite{Javier14}, \cite{Arpaia2020}, \cite{Angrisani2020} \\ 
 & (total 5) \\ 
 \hline
 \textbf{Field of view} & \cite{Wittich2018} (total 1) \\ 
 \hline
\end{tabular}
\end{center}

\section{Software Mainly Used For Each HMD}\label{ThirdAppendix}

\begin{center}
\small
\begin{tabular}{ |c|c| } 
 \hline
 \textbf{VPL Eyephonc LX} & Reality Built For Two (RB2) \\ 
 \hline
 \textbf{Virtual Research VR 4} & Developers created software \\ 
 \hline
 \textbf{i-glasses} & ViO SDK, Win95 driver \\ 
 \hline
 \textbf{Visette Pro} & Computer-based \\ 
 & software created applications \\ 
 \hline
 \textbf{NVIS nVisor} & Computer modeling software “GO” \\ 
 \hline
 \textbf{Google glass} & Glass Development Kit \\ 
 \hline
 \textbf{Oculus} & Oculus Developer Hub (ODH), \\ 
 & Unreal Engine, Unity \\ 
 \hline
 \textbf{Moverio BT-2000} & Android platform control \\ 
 & by EPSON own API \\ 
 \hline
 \textbf{HTC Vive} & VIVE WAVE, VIVEPORT, \\ 
 & VIVE Sense, OpenXR \\ 
 \hline
 \textbf{Meta 2} & Meta 2 SDK, Unity \\ 
 \hline
 \textbf{Samsung Gear VR} & Unity \\ 
 \hline
 \textbf{Microsoft HoloLens} & Mixed Reality ToolKit, Unity, \\ 
 & Unreal Engine, OpenXR, JavaScript \\ 
 \hline
\end{tabular}
\end{center}

\section*{Acknowledgments}
This research was sponsored by the DEVCOM U.S. Army Research Laboratory under Cooperative Agreement Number W911NF-21-2-0145 to B.P. 
\\
We thank Jessica Schultheis for very helpful editing. Any errors or omission are those of the authors. 
\\
The views and conclusions contained in this document are those of the authors and should not be interpreted as representing the official policies, either expressed or implied, of the DEVCOM Army Research Laboratory or the U.S. Government. The U.S. Government is authorized to reproduce and distribute reprints for Government purposes notwithstanding any copyright notation.

\bibliographystyle{IEEEtran}
\bibliography{AR_MR_Survey}

\end{document}